\documentstyle[11pt]{article}
\input amssymb.sty

\title{Representing Quantum Superpositions:\\Powers, Potentia and Potential Effectuations}

\author{{\sc Christian de Ronde}}
\date{}

\begin{document}

\bibliographystyle{plain}
\maketitle

\begin{center}
\begin{small}
Instituto de Filosof\'ia ``Dr. A. Korn" \\ 
Universidad de Buenos Aires, CONICET - Argentina \\
Center Leo Apostel and Foundations of  the Exact Sciences\\
Brussels Free University - Belgium \\
\end{small}
\end{center}

\begin{abstract}
\noindent In this paper we attempt to provide a physical representation of quantum superpositions. For this purpose we discuss the constraints of the quantum formalism to the notion of {\it possibility} and the necessity to consider a potential realm independent of actuality. Taking these insights into account and from the basic principles of quantum mechanics itself we advance towards the definition of the notions of {\it power} and {\it potentia}. Assuming these notions as a standpoint we analyze the meaning of `observation' and `interaction'. As a conclusion we provide a set of axioms which comprise our interpretation of quantum mechanics and argue in favor of a redefinition of the orthodox problems discussed in the literature. 
\end{abstract}
\begin{small}


\end{small}

\bibliography{pom}

\begin{thebibliography}{10}

\bibitem{Aerts10} Aerts, D., 2010, ``A Potentiality and
Conceptuality Interpretation of Quantum Mechanics'', {\it
Philosophica}, {\bf 83}, 15-52.

\bibitem{Aristotle} Aristotle, 1995, {\it The Complete Works of Aristotle}, The Revised Oxford
Translation, J. Barnes (Ed.), Princeton University Press,
New Jersey.

\bibitem{BeneDieks02} Bene, G. and Dieks, D., 2002, ``A Perspectival Version of the Modal Interpretation of Quantum Mechanics and the
Origin of Macroscopic Behavior'', {\it Foundations of Physics}, {\bf 32}, 645-671.

\bibitem{Nature13} Bernien, H., Hensen, B., Pfaff, W., Koolstra, G., Blok, M. S., Robledo, L., Taminiau, T. H., Markham, M., Twitchen, D. J., Childress, L. and Hanson, R., 2013, ``Heralded entanglement between solid-state qubits separated by three metres'',  {\it Nature}, {\bf 497}, 86-90.

\bibitem{Bohr34} Bohr, N., 1934, {\it Atomic theory and the
description of nature}, Cambridge University Press, Cambridge.

\bibitem{BokulichBokulich05} Bokulich, P. and Bokulich, A., 2005,
``Niels Bohr's Generalization of Classical Mechanics'', {\it
Foundations of Physics}, {\bf 35}, 347-371.

\bibitem{Cartwright78} Cartwright, N., 1978, ``The Only Real Probabilities in Quantum Mechanics'', in {\it PSA: Proceedings of the Biennial Meeting of the Philosophy of Science Association}, 54-59, The University of Chicago Press.

\bibitem{Nature11a} Clausen, C., Usmani, I., Bussi\`eres, F., Sangouard, N., Afzelius, M., de Riedmatten, H. and Gisin, N., 2011, ``Quantum storage of photonic entanglement in a crystal'',  {\it Nature}, {\bf 469}, 508-511.

\bibitem{daCostadeRonde13} da Costa, N. and de Ronde, C., 2013, ``The Paraconsistent Logic of Quantum Superpositions", {\it Foundations of Physics}, {\bf 43}, 845-858.

\bibitem{deRonde10} de Ronde, C., 2010, ``For and Against Metaphysics
in the Modal Interpretation of Quantum Mechanics", {\it
Philosophica}, {\bf 83}, 85-117.

\bibitem{deRonde11} de Ronde, C., 2011, {\it The Contextual and Modal Character of Quantum
Mechanics: A Formal and Philosophical Analysis in the Foundations of
Physics}, PhD dissertation, Utrecht University.

\bibitem{deRonde11b} de Ronde, C., 2011, ``La noci\'{o}n de potencialidad ontol\'{o}gica en la
interpretaci\'{o}n modal de la mec\'{a}nica cu\'{a}ntica", {\it
Scientiae Studia}, {\bf 10}, 137-164.

\bibitem{deRonde13a} de Ronde, C., 2013, ``The Problem of Representation and Experience in Quantum Mechanics'', in {\it Probing the Meaning of Quantum Mechanics: Physical, Philosophical and Logical Perspectives},  D. Aerts, S. Aerts and C. de Ronde (Eds.), World Scientific, Singapore, in press.

\bibitem{deRonde13b} de Ronde, C., 2013, ``Quantum Superpositions and Causality: On the Multiple Paths to the Measurement Result", preprint.

\bibitem{deRondeBontems} de Ronde, C. and Bontems, V., 2011, ``La notion d'entit\'{e} en tant qu'obstacle \'{e}pist\'{e}mologique: Bachelard, la m\'{e}cacique
quantique et la logique'', {\it Bulletin des Amis de Gaston Bachelard}, {\bf 13}, 12-38.

\bibitem{RFD13} de Ronde, C., Freytes, H. and Domenech, G., 2014, ``Interpreting the Modal Kochen-Specker Theorem: Possibility and Many Worlds in Quantum Mechanics'', {\it Studies in History and Philosophy of Modern Science}, {\bf 45}, 11-18.

\bibitem{RFD13b} de Ronde, C., Freytes, H. and Domenech, G., 2014, ``Quantum Mechanics and the Interpretation of the Orthomodular Square of Opposition'', {\it History and Philosophy of Logic}, sent.

\bibitem{Dieks88a} Dieks, D., 1988, ``The Formalism of Quantum
Theory: An Objective Description of Reality'', {\it Annalen der
Physik}, {\bf 7}, 174-190.

\bibitem{Dieks07} Dieks, D., 2007, ``Probability in the modal
interpretation of quantum mechanics'', {\it Studies in History and
Philosophy of Modern Physics}, {\bf 38}, 292-310.

\bibitem{DF} Domenech, G. and Freytes, H., 2005, ``Contextual
logic for quantum systems'', {\it Journal of Mathematical Physics},
{\bf 46}, 012102-1 - 012102-9.

\bibitem{DFR06} Domenech, G., Freytes, H. and de Ronde, C., 2006,
``Scopes and limits of modality in quantum mechanics",
\textit{Annalen der Physik}, {\bf 15}, 853-860.

\bibitem{DFR08} Domenech, G., Freytes, H. and de Ronde, C., 2008,
``A topological study of contextuality and modality in quantum
mechanics'', {\it International Journal of Theoretical Physics},
{\bf  47}, 168-174.

\bibitem{Dorato11} Dorato, M., 2011, ``Do Dispositions and
Propensities have a role in the Ontology of Quantum Mechanics? Some
Critical Remarks", in {\it Probabilities, Causes, and Propensities
in Physics}, 197-218, M. Su\'arez (Ed.), Synthese Library, Springer,
Dordrecht.

\bibitem{DoratoEsfeld} Dorato, M. and Esfeld, M., ``GRW as an Ontology of Dispositions'', {\it Studies in History and Philosophy of Modern Physics}, {\bf 41}, 41-49 .

\bibitem{Everett57} Everett, H., 1957, {\it On the Foundations
of Quantum Mechanics}, Doctoral dissertation, Princeton University,
Princeton.

\bibitem{Everett73} Everett, H., 1973, ``The Theory of the
Universal Wave Function'', in {\it The Many-Worlds Interpretation of
Quantum Mechanics}, DeWitt and Graham (Eds.), Princeton University
Press, Princeton.

\bibitem{Fine86} Fine, A., 1986, {\it The Shaky Game}, University of Chicago Press, Chicago. 

\bibitem{GRW} Ghirardi, G. C., Rimini A. and Weber, T., 1986, ``Unified
Dynamics for Microscopic and Macroscopic Systems'', {\it Physical
Review D}, {\bf 34}, 470-491.

\bibitem{Healey89} Healey, R., 1989, {\it The Philosophy of Quantum Mechanics: An Interactive 
Interpretation}, Cambridge University Press, Cambridge.

\bibitem{Heis58} Heisenberg, W., 1958, {\it Physics and Philosophy},
World perspectives, George Allen and Unwin Ltd., London.

\bibitem{Jauch68} Jauch, J. M., 1968, {\it Foundations of quantum
mechanics}, Addison-Wesley, MA.

\bibitem{Jaynes} Jaynes, E. T., 1990, {\it Complexity, Entropy, and the Physics
of Information}, edited by W. H. Zurek Addison-Wesley.

\bibitem{KS} Kochen, S. and Specker, E., 1967, ``On the problem
of Hidden Variables in Quantum Mechanics", {\it Journal of
Mathematics and Mechanics}, {\bf 17}, 59-87. Reprinted in Hooker,
1975, 293-328.

\bibitem{Laurikainen98} Laurikainen, K. V., 1998, {\it The Message
of the Atoms, Essays on Wolfgang Pauli and the Unspeakable}, Spinger
Verlag, Berlin.

\bibitem{NaturePhy12} Ma, X., Zotter, S., Kofler, J., Ursin, R., Jennewein, T., Brukner, C. and Zeilinger, A.,  2012, ``Experimental delayed-choice entanglement swapping'',  {\it Nature Physics}, {\bf 8}, 480-485.

\bibitem{MM} Maeda, F. and  Maeda, S., 1970,  {\it Theory
of Symmetric Lattices}, Springer-Verlag, Berlin.

\bibitem{Margenau54} Margenau, H., 1954, ``Advantages and disadvantages of various
interpretations of the quantum theory", {\it Physics Today}, {\bf
7}, 6-13.

\bibitem{Nature07} Ourjoumtsev, A., Jeong, H., Tualle-Brouri, R. and
Grangier, P., 2007, ``Generation of optical `Schr\"{o}dinger cats'
from photon number states'', {\it Nature}, {\bf 448}, 784-786.

\bibitem{PauliJung} Pauli, W. and Jung, C. G., 2001, {\it Atom and
Archetype, The Pauli/Jung Letters 1932-1958}, Princeton University
Press, New Jersey.

\bibitem{Piron76} Piron, C., 1976, {\it Foundations of Quantum
Physics}, W.A. Benjamin Inc., Massachusetts.

\bibitem{Piron99} Piron, C., 1999, ``Quanta and Relativity: Two
Failed Revolutions'', In {\it The White Book of Einstein Meets
Magritte}, 107-112, D. Aerts J. Broekaert and E. Mathijs (Eds.),
Kluwer Academic Publishers.

\bibitem{Popper82} Popper, K. R., 1982, {\it Quantum Theory and the Schism in Physics},  Rowman and Littlefield, New Jersey.

\bibitem{PBR} Pusey, M. F.,  Barrett, J. and Rudolph, T., 2012, ``On the reality of the quantum state'', {\it Nature Physics}, {\bf 8}, 475-478.

\bibitem{Albert} Sol\'e, A., 201``Bohmian mechanics without wave function ontology'', 
{\it Studies in History and Philosophy of Modern Science}, {\bf 44}, 365-378. 

\bibitem{Suarez04b} Su\'arez, M., 2004, ``Quantum Selections, Propensities, and the
Problem of Measurement'', {\it British Journal for the Philosophy of
Science}, {\bf 55}, 219-255.

\bibitem{Suarez07} Su\'arez, M., 2007, ``Quantum propensities'',
{\it Studies in History and Philosophy of Modern Physics}, {\bf 38},
418-438.

\bibitem{VF91} Van Fraassen, B. C., 1991, {\it Quantum Mechanics: An
Empiricist View}, Clarendon, Oxford.

\bibitem{VF02} Van Fraassen, B. C., 2002, {\it The Empirical Stance}, Yale University
Press, New Haven.

\bibitem{VN} Von Neumann, J., 1996, {\it Mathematical Foundations
of Quantum Mechanics}, Princeton University Press (12th. edition),
Princeton.

\bibitem{WZ} Wheeler, J. A. and Zurek, W. H. (Eds.), 1983, {\it Theory and
Measurement}, Princeton University Press, Princeton.

\end{thebibliography}

\newtheorem{theo}{Theorem}[section]

\newtheorem{definition}[theo]{Definition}

\newtheorem{lem}[theo]{Lemma}

\newtheorem{met}[theo]{Method}

\newtheorem{prop}[theo]{Proposition}

\newtheorem{coro}[theo]{Corollary}

\newtheorem{exam}[theo]{Example}

\newtheorem{rema}[theo]{Remark}{\hspace*{4mm}}

\newtheorem{example}[theo]{Example}

\newcommand{\proof}{\noindent {\em Proof:\/}{\hspace*{4mm}}}

\newcommand{\qed}{\hfill$\Box$}

\newcommand{\ninv}{\mathord{\sim}} 

\newtheorem{postulate}[theo]{Postulate}

\begin{flushright}
\emph{We are all agreed that your theory is crazy.\\
 The question that divides us is whether it is\\
 crazy enough to have a chance of being correct.}\\
\vspace{0.5cm}
Niels Bohr
\end{flushright}
\vspace{0.5cm}

\section*{Introduction}

\noindent Quantum superpositions constitute one of the main formal elements used today in laboratories around the world to produce some of the most outstanding developments in what could be called the new quantum technology. But, what is a quantum superposition? There is no consensus in the physics community about what should be the answer to this question \cite{PBR}. Quantum superpositions seem in the laboratory ontologically robust. We can use superpositions to teleport information, to implement quantum computers, but we still cannot find the physical concept which unifies all we have learnt about them. Indeed, there are many characteristics and behaviours we know about superpositions: we know about {\it their existence regardless of the effectuation of one of its terms}, as shown, for example, by the interference of different possibilities in {\it welcher-weg} type experiments \cite{ Nature11a, NaturePhy12}, {\it their reference to contradictory properties}, as in Schr\"{o}dinger cat states \cite{Nature07}, we also know about {\it their non-standard route to actuality}, as explicitly shown by the MKS theorem \cite{DFR06, RFD13}, and we even know about {\it their non-classical interference with themselves and with other superpositions}, used today within the latest technical developments in quantum information processing \cite{Nature13}. However, we still cannot say what a quantum superposition {\it is} or {\it represents}. 

The fact is that since its origin QM has confronted us with the limits of our classical representation of the world. For many, we need to give up on visualizability, abandon representation and content ourselves with a mathematical formalism which predicts probabilistically the correct measurement outcomes for a given experimental arrangement. As remarked by Arthur Fine: 

\begin{quotation}
{\small ``[The] instrumentalist moves, away from a realist construal of the emerging quantum theory, were given particular force by Bohr's so-called Ôphilosophy of complementarityÕ; and this non-realist position was consolidated at the time of the famous Solvay conference, in October of 1927, and is firmly in place today. Such quantum non-realism is part of what every graduate physicist learns and practices. It is the conceptual backdrop to all the brilliant successes in atomic, nuclear, and particle physics over the past fifty years. Physicists have learned to think about their theory in a highly non-realist way, and doing just that has brought about the most marvelous predictive success in the history of science.'' \cite[p. 88]{Fine86}}
\end{quotation}

\noindent For others, this answer is unacceptable, for physics is a discipline which seeks to understand and describe the world and nature. From this perspective, one is forced to provide an answer to the question: what is physical reality according to QM? Specially after the triumph of Niels Bohr, which allowed physics to evade metaphysical issues and a representational construal of the theory, this question faces serious difficulties.

\begin{quotation}
{\small ``[O]ur present [quantum mechanical] formalism is not purely epistemological; it is a peculiar mixture describing in part realities of Nature, in part incomplete human information about Nature ---all scrambled up by Heisenberg and Bohr into an omelette that nobody has seen how to unscramble. Yet we think that the unscrambling is a prerequisite for any further advance in basic physical theory. For, if we cannot separate the subjective and objective aspects of the formalism, we cannot know what we are talking about; it is just that simple.'' \cite[p. 381]{Jaynes}}
\end{quotation}

Many attempts have been made, but none ---due to the many problems found within each of the very different proposals--- has been able to come up with a coherent solution to the question of interpretation in QM. Elsewhere \cite{deRonde10}, we have argued that one can find in the vast literature regarding the interpretation of QM, two main strategies which attempt to provide an answer to the riddle of what QM is talking about. The first strategy is to begin with a presupposed set of metaphysical principles and advance towards a new formalism. Examples of this strategy are Bohmian mechanics, which presupposes the existence of positions and trajectories, and the collapse theory proposed by Ghirardi, Rimini and Weber (also called `GRW theory') \cite{GRW}, which introduces non-linear terms in the Schr\"odinger equation. But these proposals, going beyond the orthodox formalism, rather than interpreting QM seem to create new theories ---each of which can even possess multiple interpretations \cite{Albert}. Following Healey, we agree that this is not discussing the interpretation of QM, but rather ``changing the subject'' \cite[p. 24]{Healey89}. The second strategy is to accept the orthodox formalism of QM and advance towards the creation and elucidation of new metaphysical principles which match the mathematical structure. Examples of this second strategy are quantum logic and its different lines of development, such as the Geneva School of Jauch and Piron \cite{Jauch68, Piron76}, the modal interpretation of van Fraassen and Dieks \cite{VF91, Dieks88a, Dieks07}, and Everett's relative state interpretation \cite{Everett57, Everett73}. From this perspective, the importance is to focus in the formalism of the theory and try to learn about the symmetries, the logical features and structural relations. The idea is that by learning about such aspects of the theory we can also develop the metaphysical conditions which should be taken into account in a coherent ontological interpretation of QM. In Everett's words: let QM find its own interpretation. 
  
Within the second strategy, there is also another distinction which can help us to specify even more the interpretational map of QM. This distinction can be made by taking into account the different solutions provided to the so called measurement problem.\footnote{In the ortodox interpretation of QM we assume that a quantum state is given by a superposition $( \Psi =  \Sigma  \ c_i | \alpha_i \rangle)$ that interacts with an apparatus ``ready to measure''   $(| R_o \rangle)$. As a result of the quantum interaction both states become entangled  $( \Sigma  \ c_i | \alpha_i \rangle | R_o \rangle \rightarrow \Sigma  \ c_i | \alpha_i \rangle | R_i \rangle)$. However, when a measurement is produced we observe in the apparatus a single result $(| R_k \rangle)$. The measurement problem discusses the justification of the path from a quantum superposition to one single result, instead of a superposition of multiple results.} On the one hand, there is a ``collapse solution'' which ---going back to Heisenberg, Popper and Margenau \cite{Heis58, Popper82, Margenau54}--- attempts to distinguish a realm, different to actuality, which contains potentialities, propensities or latencies.\footnote{This path has been subject of research during the last decades by authors like Diederik Aerts \cite{Aerts10}, who continues the line of research of the Geneva School, and by Mauricio Su\'arez, Mauro Dorato and Michael Esfeld \cite{Dorato11, DoratoEsfeld, Suarez04b, Suarez07}, who have continued developing in different ways the propensity interpretation of Popper.} According to this view, during the process of measurement there is a ``collapse'' (i.e., a real physical interaction) which takes one of the possible terms in the quantum superposition into actuality. On the other hand, there is an ``actualist'' or  ``non-collapse solution'' which attempts to ``restore a classical way of thinking about {\it what there is}.'' This proposal interprets, in very different ways, the quantum superposition as describing one or many actual properties. For example, while many worlds interpretation considers that all terms in the superposition are actual,\footnote{As remarked by Hugh Everett \cite[p. 146-147]{Everett73} himself: ``The whole issue of the transition from `possible' to `actual' is taken care of in the theory in a very simple way ---there is no such transition, nor is such a transition necessary for the theory to be in accord with our experience. From the viewpoint of the theory all elements of a superposition (all `branches') are `actual', none any more `real' than the rest.''} in the modal interpretation of Dieks only one of the properties is actual (the one in which the pure state is written as a single term) and the rest are considered as possible properties \cite{Dieks07}. In these ``non-collapse'' solutions `possibility' is regarded, in the same way it is done in classical physics, as an `epistemic possibility'. However, as different as they might seem at first sight, we have argued that both lines of research ---``collapse'' and ``non-collapse'' solutions--- concentrate their efforts in justifying the actual realm of existence \cite{deRonde13b}. 

Following Wolfgang Pauli, we understand that QM confronts us with a redefinition of our idea of reality.\footnote{As remarked by Pauli \cite[p. 193]{Laurikainen98}: ``When the layman says `reality' he usually thinks that he is speaking about something which is self-evidently known; while to me it appears to be specifically the most important and extremely difficult task of our time to work on the elaboration of a new idea of reality.''} Our research has been focused in providing a physical representation of QM which, starting from the formalism itself, provides a new understanding of reality, an understanding which avoids the implicit equation: actuality = reality. Our strategy has concentrated, on the one hand, in arguing against the restricted actualist conception of reality,\footnote{Notice that even the teleological hylomorphic scheme falls pray of actualism ---understood as a total or partial preeminence of the actual mode of existence--- since the notion of potentiality restricts itself to {\it becoming actual}. We will come back to this point in sec. 2.} and on the other, in developing an idea of {\it potential reality} which is truly independent of the actual realm \cite{deRonde11}.

\section{Formal Constraints on Quantum Possibility}

Contextuality can be directly related to the impossibility to represent a piece of the world as constituted by a set of definite valued properties independently of the choice of the context. In formal terms, this is demonstrated by the Kochen-Specker (KS) theorem \cite{KS}, which only makes reference to the actual realm. But as we know, QM makes probabilistic assertions about measurement results. Therefore, it seems natural to assume that QM does not only deal with actualities but also with possibilities. Then the question arises whether the space of possibilities is subject to the same restrictions as the space of actualities. Given an adequate definition of the possibility operator $\Diamond$ ---as the one developed in \cite{DFR06, DFR08}--- the set of possibilities is the center of an enlarged structure. Since the elements of the center of a structure are those which commute with all other elements, one might think that the possible propositions defined in this way escape from the constraints arising from the non-commutative character of the algebra of operators. Thus, at first sight one might assume that possibilities behave in a classical manner. In order to see that this is not the case we need to review some basic notions. 

We denote by ${\cal OML}$ the variety of orthomodular lattices. Le $\mathcal{L}$ be an orthomodular lattice. {\it Boolean algebras} are orthomodular lattices satisfying  the {\it distributive law} $x\land (y \lor z) = (x \land y) \lor (x \land z)$. We denote by ${\bf 2}$
the Boolean algebra of two elements. We denote by $Z(\mathcal{L})$ the set of all central elements of $\mathcal{L}$ and it is called the {\it center} of $\mathcal{L}$.
In the tradition of quantum logical research, a property of (or
a proposition about) a quantum system is related to a closed
subspace of the Hilbert space ${\mathcal H}$ of its (pure) states
or, analogously, to the projector operator onto that subspace.
Moreover, each projector is associated to a dichotomic question
about the actuality of the property \cite[p. 247]{VN}. A physical
magnitude ${\mathcal M}$ is represented by an operator ${\bf M}$
acting over the state space. For bounded self-adjoint operators,
conditions for the existence of the spectral decomposition ${\bf
M}=\sum_{i} a_i {\bf P}_i=\sum_{i} a_i |a_i\rangle\langle a_i|$ are
satisfied. The real numbers $a_i$ are related to the outcomes of
measurements of the magnitude ${\mathcal M}$ and projectors
$|a_i\rangle\langle a_i|$ to the mentioned properties. The physical
properties of the system are organized in the lattice of closed
subspaces ${\mathcal L}({\mathcal H})$ that, for the finite
dimensional case, is a modular lattice, and an orthomodular one in
the infinite case \cite{MM}. Moreover, each self-adjoint operator
$\bf M$ has an associated Boolean sublattice $W_{\bf{M}}$ of
${\mathcal L}({\mathcal H})$ which we will refer to as the spectral
algebra of the operator $\bf M$. Assigning values to a physical
quantity ${\cal M}$ is equivalent to establishing a Boolean
homomorphism $v: W_{\bf{M}} \rightarrow {\bf 2}$. As it is well
known, the KS theorem rules out the non-contextual assignment of
definite values to the physical properties of a quantum system. This
may be expressed in terms of valuations over  ${\mathcal
L}({\mathcal H})$ in the following manner. We first introduce the
concept of global valuation. Let  $(W_i)_{i\in I}$ be the family of
Boolean sublattices of ${\mathcal L}({\mathcal H})$. Then a {\it
global valuation} of the physical magnitudes over ${\mathcal
L}({\mathcal H})$ is a family of Boolean homomorphisms $(v_i: W_i
\rightarrow {\bf 2})_{i\in I}$ such that $v_i\mid W_i \cap W_j =
v_j\mid W_i \cap W_j$ for each $i,j \in I$. If this global valuation
existed, it would allow to give values to all magnitudes at the same
time maintaining a {\it compatibility condition} in the sense that
whenever two magnitudes shear one or more projectors, the values
assigned to those projectors are the same from every context. The KS
theorem, in the algebraic terms, rules out the existence of global
valuations when $dim({\mathcal H})>2$:

\begin{theo}\label{CS2} {\rm \cite[Theorem 3.2]{DF}}
If $\mathcal{H}$ is a Hilbert space such that $dim({\cal H}) > 2$,
then a global valuation, i.e. a family of Boolean homomorphisms over
the spectral algebras satisfying the compatibility condition, over
$L({\mathcal H})$ is not possible. \qed
\end{theo}

In  \cite{DFR06, DFR08} we delineated a modal extension for orthomodular
lattices that allows to formally represent, within the same
algebraic structure, actual and possible properties of the system.
This allows us to  discuss the restrictions posed by the theory
itself to the {\it actualization} of  possible properties. Given a
proposition about the system, it is possible to define a context
from which one can predicate with certainty about it together with a
set of propositions that are compatible with it and, at the same
time, predicate probabilities about the other ones (Born rule). In
other words, one may predicate truth or falsity of all possibilities
at the same time, i.e., possibilities allow an interpretation in a
Boolean algebra. In rigorous terms, let $P$ be a proposition about a
system and consider it as an element of an orthomodular lattice
${\cal L}$. If we refer with $\Diamond P$  to the possibility of $P$
then we can assume that $\Diamond P \in
Z({\cal L})$. 

This interpretation of possibility in terms of the Boolean algebra
of central elements of ${\cal L}$ reflects the fact that one can
simultaneously predicate about all possibilities because Boolean
homomorphisms of the form $v:Z({\cal L}) \rightarrow {\bf 2}$ can be
always established. If $P$ is a proposition about the system and $P$
occurs, then it is trivially possible that $P$ occurs. This is
expressed as $P \leq \Diamond P$. Classical consequences that are
compatible with a given property, for example probability
assignments to the actuality of other propositions, share the
classical frame. These consequences are the same ones as those which
would be obtained by considering the original actual property as a
possible property. This is interpreted as, if $P$ is a property of
the system, $\Diamond P$ is the smallest central element greater
than $P$. This enriched orthomodular structure can be axiomatized as a variety denoted by ${\cal OML}^\Diamond$ \cite[Theorem 4.5]{DFR06}. 

The possibility space represents the modal content added to the
discourse about properties of the system. Within this frame, the
actualization of a possible property acquires a rigorous meaning.
Let ${\cal L}$ be an orthomodular lattice, $(W_i)_{i \in I}$ the
family of Boolean sublattices of ${\cal L}$ and ${\cal L}^\Diamond$
a modal extension of $\cal L$. If $f: \Diamond {\cal L} \rightarrow
{\bf 2}$ is a Boolean homomorphism, an actualization compatible with
$f$ is a global valuation $(v_i: W_i \rightarrow {\bf 2})_{i\in I}$
such that $v_i\mid W_i \cap \Diamond {\cal L} = f\mid W_i \cap
\Diamond {\cal L} $ for each $i\in I$. A kind of converse of this
possibility of actualizing properties may be read as an algebraic
representation of the Born rule, something that has no place in the
orthomodular lattice alone. {\it Compatible actualizations}
represent the passage from possibility to actuality, they may be
regarded as formal constrains when applying the interpretational
rules proposed in the different modal versions. When taking into
account compatible actualizations from different contexts, an
analogon of the KS theorem holds for possible properties.

\begin{theo}\label{ksm}{\rm \cite[Theorem 6.2]{DFR06}}
Let $\cal L$ be an orthomodular lattice. Then $\cal L$ admits a
global valuation iff for each possibility space there exists a
Boolean homomorphism  $f: \Diamond {\cal L} \rightarrow {\bf 2}$
that admits  a compatible actualization.\qed
\end{theo}

\noindent The MKS theorem shows that no enrichment of the
orthomodular lattice with modal propositions allows us to circumvent
the contextual character of the quantum language. Thus, from a
formal perspective, one is forced to conclude that quantum
possibility is something different from classical possibility. 

In \cite{RFD13} we were able to provide a physical interpretation of the MKS theorem relating the formal elements of the different structures with a specific set of concepts. 

\begin{enumerate}
\item $\Diamond A_i$ with $A_i$ in the Boolean lattice $\wp(\Gamma)$ is called ``possibility of $A_i$''.
\item $\Diamond_{Q} {\bf P}_i$ with ${\bf P}_i$ in the orthomodular lattice $\mathcal{L} $ is called ``quantum possibility of ${\bf P}_i$''. 
\item The set of all the  $\Diamond_{Q} {\bf
P}_i$ with ${\bf P}_i$ in the orthomodular lattice $\mathcal{L} $ is
called the ``set of quantum possibilities''.
\item A Boolean sub-algebra of the orthomodular lattice $\mathcal{L} $ is called a ``context''.
\item The set of quantum possibilities valuated to $1\in {\bf 2}$ is called the ``set of existent quantum possibilities''.
\item The subset of quantum possibilities in direct relation to a context valuated to $1\in {\bf 2}$  is
called the ``set of existent quantum possibilities in a situation''.
\item The subset of projectors of the context  valuated to $1\in {\bf 2}$ is called the ``actual state of
affairs''.
\end{enumerate}

\noindent Physically, it follows from the given definitions that:

\begin{enumerate}
\item An ``actual state of affairs''
provides a physical description in terms of definite valued
properties.
\item A ``situation'' provides a physical description in terms
of the quantum possibilities that relate to an actual state of
affairs.
\item Formally, to go from the ``set of existent quantum possibilities''  to one of its
subsets (each of which relates to a ``context'') is to define an
application; physically, this path relates to the choice of a
particular measurement set up, restricting the expressiveness of the total set of existent possibilities to a specific subset.
\item Formally, to give values to the projectors ${\bf
P}_i$ in a context is to valuate; physically, the valuation
determines the set of properties (in correspondence with the
projectors ${\bf P}_i$ valuated to $1\in {\bf 2}$) which are
considered as preexistent.
\item The ``situation'' expresses an existent set of quantum possibilities
(which must not be considered in terms of actuality) while the
valuated context  expresses an actual state of affairs. This leaves
open the opportunity to consider quantum possibility as determining
a different mode of existence (independent of that of actuality).
\end{enumerate}

\noindent Indeed, this analysis shows that the realm of possibility must be considered ---at least formally--- as independent to the actual realm. Forcing the classical notion of possibility within the quantum structure is a move that contradicts the mathematical formalism, which contemplates the interaction of possible existents similarly as how classical physics contemplates the interaction of actual existents.

\section{Potential State of Affairs}

Aristotle articulated his metaphysics in terms of a teleological hylomorphic scheme. As remarked by Pauli \cite[p. 93]{PauliJung},  ``Aristotle created the important concept of {\it potential being} and applied it to {\it hyle}'', but ``he was not able to fully carry out his intention to grasp the \emph{potential}, and his endeavors became bogged down in early stages.'' Indeed, his project to consider different modes of being ---i.e., the potential and the actual--- was betrayed when he himself choose to center the fundament of his architectonic in {\it pure acto}. Potentiality was then understood only in terms of actuality, more specifically, the potential was understood as `lack', `imperfect' and `incomplete' with respect to the actual. As remarked by Aristotle in his {\it Metaphysics}: ``We have distinguished the various senses of `prior', and it is clear that actuality is prior to potentiality. [...] For the action is the end, and the actuality is the action. Therefore even the word `actuality' is derived from `action', and points to the fulfillment'' [1050a17-1050a23]. Aristotle then continues to provide arguments in this line which show ``[t]hat the good actuality is better and more valuable than the good potentiality'' [1051a4-1051a17]. More than two millennia later, 17th century metaphysics prepared, through the division of {\it res cogitans} and {\it res extensa}, the condotions of possibility for Newton to eliminate completely the potential realm in physics ---doing also away with the {\it final cause}. His mechanics became a physics of pure actuality, a physics which provided a description of the universe in terms of an {\it Actual State of Affairs} (ASA), supplemented by the {\it efficient cause}. Only with the adveniment of QM ---the principle of indetermination and the ``infamous quantum jumps''--- the ground was ready for Heisenberg to recover the potential realm, but once again, only for the purpose of explaining the actual \cite{RFD13}. As remarked by Heisenberg: 

\begin{quotation}
{\small ``I believe that the language actually used by physicists
when they speak about atomic events produces in their minds similar
notions as the concept of `potentia'. So physicists have gradually
become accustomed to considering the electronic orbits, etc., {\it not as
reality} but rather as a kind of `potentia'.'' \cite[p. 156, my emphasis]{Heis58}}
\end{quotation}

Our research has analyzed the idea of considering a mode of existence truly independent of actuality. According to our proposal one can advance towards such an independent notion provided we develop a scheme which goes beyond actuality. 
In order to advance in this direction  we need to introduce the idea of a {\it Potential State of Affairs} (PSA) ---what we have called in \cite{RFD13} `set of existent quantum possibilities'---, the notion of potential effectuation and the immanent cause \cite{deRonde13b}. Indeed, by using these new concepts, we expect that our previous formal analysis regarding quantum possibility can find a suitable physical interpretation. But now the question arises: what are the ``things'' which exist and interact within this potential realm? Our answer is: {\it powers} with definite {\it potentia}.

\section{A World Made of Powers}

The history of modern physics can be described as the history of physical entities. Waves, fields, bodies, etc., they all follow the logical and ontological principles of the notion of `entity'. But concepts are not God given, they are creations, so just like the concept of entity was constructed by the human intellect, it is in principle possible to think in a completely different concept which could allow us to talk about physical reality. Our research has focused on this problem: how can we develop a set of concepts ---unscrambling the omelette of Heisenberg and Bohr--- which bring into stage that of which QM is talking about in terms of an objective account of physical reality? The radical departure of our approach with respect to the classical understanding of reality is given by the fact we are ready to give up on the exclusiveness of the actual realm of existence and explore the original dictum of Aristotle, that {\it being is said in different ways.}

The first important point according to our stance is to recall the fact that Aristotle grounded the notion of entity in the logical and ontological {\bf principles of existence}, {\bf non-contradiction} and {\bf identity}. Our proposal is that in fact there exist analogous principles in QM which can allow us to develop new concepts. The {\bf principles of indetermination}, {\bf superposition} and {\bf difference} could be considered as providing the logical and ontological foundation of that of which QM is talking about. As we shall see these principles do not lead to the notion of `entity'. So which are the concepts that match these principles in a coherent manner?

An experimental arrangement is nothing but the condition of possibility for an action to take place, it creates the capability to perform an experiment. In QM we always talk about such experimental arrangements and the possible outcomes they expose. In this respect, we are also faced with the {\it choice} of mutually incompatible experimental arrangements, each of which expresses a given set of capabilities. And it is here that the scrambling of both objectivity and subjectivity threatens, for as we know the KS theorem shows us that all possible experimental arrangements cannot be conceived as making reference to the same physical object nor an ASA. Subjectivity only enters the scene when we try to impose the notion of entity within this formal structure. It is at this point that we have criticized the notion of entity as an epistemological obstruction \cite{deRondeBontems}. At the same time, in order to escape the limits of the ontology of entities, which exist in the mode of being of actuality, we have investigated the possibility to put forward an ontology of powers which exist in the mode of being of ontological potentiality \cite{deRonde11, deRonde11b}. We claim that just like the logical and ontological {\bf principles of existence}, {\bf non-contradiction} and {\bf identity} provide the constraints for a proper understanding of the concept of {\it entity}; the {\bf principles of indetermination}, {\bf superposition} and {\bf difference} are able to determine the notion of {\it power}.

\subsection{Indetermination (Instead of Determination)}

Powers are formally represented by vectors in Hilbert space. In order to understand what we mean by a power one has to keep in mind two general rules which are not so easy to follow. Firstly, we have to forget about a direct reference to entities, and even though already language forces us into this ``Socratic trap'', we should avoid from now on committing ourselves to this particular metaphysical view. Secondly, in order to avoid thinking in the old terms of (classical) potentiality, in terms of a dynamical process, in terms of logical possibility, we should always think of powers as existents in the present tense, as elements which exist {\it hic et nunc} independently of what can actually be the case. With these two ideas in mind we are now ready to continue.

{\it Powers are indetermined.} Powers are a conceptual machinery which can allow us to compress the quantum experience into a picture of the world, just like entities such as particles, waves and fields, allow us to do so in classical physics. We cannot ``see" powers in the same way we see objects.\footnote{It is important to notice there is no difference in this point with the case of entities: we cannot ``see" entities ---not in the sense of having a complete access to them. We only see perspectives which are unified through the notion of object.} Powers are experienced in actuality through {\it elementary procesess}. A power is sustained by a logic of actions which do not necessarily take place, it \emph{is} and \emph{is not}, {\it hic et nunc}. Heisenberg's indeterminacy principle must be understood in this case as providing a mathematical expression of this basic character of powers which refers to its being {\it indetermined}.

{\it The mode of being of a power is potentiality}, not thought in terms of classical possibility (which relies on actuality) but rather as a mode of existence ---i.e., in terms of ontological potentiality. To possess the power of {\it raising my hand}, does not mean that in the future `I {\it will} raise my hand' {\it or} that in the future `I {\it will not} raise my hand'; what it means is that, here and now, I possess a power which exists in the mode of being of potentiality, independently of what will happen in actuality. Powers do not exist in the mode of being of actuality, they are not actual existents, they are undetermined potential existents. Powers, like entities, preexist to observation, unlike entities, preexistence is not defined in the actual mode of being as an ASA, instead we have a {\it potential preexistence}, or in other words, a PSA.

\subsection{Superposition (Instead of Non-Contradiction)}

A basic question which we have posed to ourselves regards the ontological meaning of a {\it quantum superposition} \cite{deRonde11}. What does it mean to have a mathematical expression such as: $\alpha \ | \uparrow \rangle + \beta \ |\downarrow\rangle$, which allows us to predict precisely, in probabilistic terms, definite experimental outcomes? Our theory of powers has been explicitly developed in order to try to answer this particular question.

{\it Powers can be superposed to different ---even contradictory--- powers.} We understand a quantum superposition as encoding a set of powers each of which possesses a definite {\it potentia}. This we call a {\it faculty}. For example, the faculty represented by the superposition $\alpha \ | \uparrow \rangle + \beta \ |\downarrow\rangle$, combines the contradictory powers, $| \uparrow \rangle$ and $|\downarrow\rangle$, with their potentia, $\alpha$ and $\beta$, respectively. Contrary to the orthodox interpretation of the quantum state, we do not assume the identity of the multiple mathematical representations given by the different basis. Each superposition is basis dependent and must be considered as a distinct faculty. For example, the faculties $\frac{1}{\sqrt{2}} [ | \uparrow_{x} \rangle + |\downarrow_{x}\rangle]$ and $\frac{1}{\sqrt{2}} [ | \uparrow_{y} \rangle + |\downarrow_{y}\rangle]$, which can be derived from one another via a change in basis, are considered as two different and distinct faculties, $F_{\uparrow\downarrow_{x}}$ and $F_{\uparrow\downarrow_{y}}$. 

The logical structure of a superposition is such that a power and its opposite can exist at one and the same time, violating the principle of non-contradiction \cite{daCostadeRonde13}. Within the faculty of raising my hand, both powers (i.e., the power `I am {\it able to} raise my hand' and the power `I am {\it able not to} raise my hand') co-exist in the definition itself of the faculty. The faculty is {\it compressed activity}, something which {\it is} and {\it is not} the case, {\it hic et nunc}. It cannot be thought in terms of identity but is expressed as a difference, as a {\it quantum}.

\subsection{Difference (Instead of Identity)}

{\it A power is experienced in actuality as a difference.} Our concept of power confronts a problem which does not find an answer in terms of entities. How can we think of something which is {\it different} every time it is realized in an experimental procedure but rests simultaneously `the same' each time we experience it? This is the tension found in the quantum wave function which comprises simultaneously a `statistical character', since it provides the average statistical value of observables, and an `individual character', since it relates to an individual imprint (measurement outcome). If a superposition were some kind of entity it would be destroyed each time an experimental result would be expressed, so what would allow us to think it is more than its mere collapse? The paradox is then set, there is no possibility to unify the multiple actual effectuations into the identity of an entity. It is the notion of power which, supplemented by the immanent cause, allows us to unify the multiplicity of difference. Let us use the following example in order to clarify our proposal: I throw a stone in a lake. I wait a few seconds and I throw a second stone in more or less the same place. I then repeat this procedure for several hours. How can we analyze this situation? What is `the same'? What can be learnt from the multiple actual effectuations given by each one of the stones being thrown? In the first place, with just one throw, we learn about the existence of the power of throwing stones. In the second place, if we measure the distance between the average spot in which the stones are thrown and the shore, we can also learn about the potentia of this power. An important point is that this power exists independently of the fact it is effectuated in actuality, it even exists in the case there is no lake nor stones. But what is `the same'? The stones are not  the same ---each one of the stones is different---, neither the lake, which has with each throw more stones than before. The only thing which remains the same each time I choose to throw a stone is the action itself, the process and its actual effectuation. It is through an elementary process that a power is exposed in actuality. Behind the process and its effectuation stands the power, just in the same way when we observe a cup or a table the multiple perspectives are brought into one by the notion of object. It is the power, expressed in actuality through multiple elementary effectuations that of which QM is talking about. 

{\it A power has no identity.} `Sameness' changes its meaning, it is not anymore putting together that which is an identity, but rather that which is different.  If thought in terms of faculties, the classical notion of {\it identity} simply losses its meaning. Individuality is then given by the {\it causal unity} which comprises the multiplicity of elementary processes. A power is not a substantive of which one can predicate certain properties. A power is a verb, pure activity, and it makes no sense to talk of verbs as having identity. It makes no sense to ask wether the power of throwing stones is one and the same through time. This is simply a badly posed question, or at least, one that cannot be made presupposing the classical logical principle of identity. One can make this question with respect to entities because entities exist as {\it essences}, and in this case there is something which remains the same and equal to itself, but a power is not an entity, it is not an identity, a power is each time different to itself.

The orthodox interpretation of QM considers the superposition as describing the state of a quantum particle, more specifically, as encoding the probability of finding a definite property of the quantum state. In our interpretation the superposition encodes powers and potentia exposed in actuality through {\it elementary processes}. More specifically, the {\it power} is the capability to express through an elementary process an actual effectuation, while the {\it potentia} is the strength of the power given a PSA. The potentia of a given power in a definite PSA is given through the Born rule. Thus, given a definite PSA, the powers and their potentia are univocally determined. This can be seen from the fact that in order to calculate the potentia of a given power we only need the PSA and there is no need to identify the specific context of inquiry.

\begin{equation}
Potentia \ (| \alpha_{i} \rangle, \Psi) = \langle \Psi | P_{\alpha_{i}} | \Psi \rangle
\end{equation}

\subsection{No Space, No Time}

The issue of space and time has been a matter of great debate in the whole history of QM. Newton created the concepts of absolute space and time in order to reach the highest peak of a physics of pure actuality, a physics in which the realm of potentiality was completely neglected. As we argued in \cite{deRonde11}, we believe that the positivist fight of Ernst Mach against the dogmatic understanding of Newtonian absolute space and time produced the conditions of possibility for the creation of both QM and Relativity Theory. This crisis in the foundations of our classical understanding of the world produced the soil capable of developing formalisms which went beyond the preconditions of our classical experience. 

{\it Powers are non-spatial and non-temporal}. Within our interpretation the issue is to be resolved ``right from the start'': powers do not inhabit space nor time. A power cannot be thought as existing in space-time. It is only the {\it process}, through which the power is exposed, that space-time enters the scene. The process builds a bridge to bring the power from its potential existence into its space-time actual effectuation. This is the way through which a power is exposed, it is through an elementary process that the power makes contact with the actual.

\subsection{The Contextual Nature of Powers}

In orthodox QM one starts from the quantum state of the system ---forcing implicitly the notion of entity within the formalism--- and finds out later that its properties cannot be thought as preexistent ---i.e., determined independently of the measurement set up. As we discussed above, quantum contextuality makes reference to the impossibility to represent the orthodox formal structure of the theory in terms of an ASA. Only once a choice is made to single out a particular context, by explicitly neglecting the rest of the contexts, we seem to recover a classical Boolean structure for the chosen one. As we have argued in \cite{RFD13} this interpretational move does not grant that classicality is regained. It is only by neglecting counterfactual reasoning between the multiple contexts that one can do {\it as if} the chosen context reflected an ASA\cite{BeneDieks02}. But it is counterfactual reasoning itself which allows to think, even in one single context, about a system or, more generally, an ASA. The mess becomes even worse once we realize that our knowledge relates to all observables pertaining to all different contexts (i.e., the PSA) simultaneously, quite independently of the fact we choose to actualize a specific context through the construction of an experimental arrangement in the laboratory or not. As a matter of fact, given a PSA the probabilities of each one of the observables are univocally determined, or in other words, given a PSA, we can define all the existent powers and their respective potentia without actually performing any experiment.  

{\it A power is a relational existent}. A power is not composed by essences but by relations. It exists only in relation to other powers (in a definite PSA). It does not exist as a classical property independently of other properties. It exists as being indetermined, relating existence to a capability of being. One might possess a power, but in order to actually effectuate it there are specific conditions that need to become actual (e.g., in order to throw stones in a lake I need to have stones and a lake, in order to swim I need a swimming pool, in order to play my guitar I need my guitar, etc.). Thus, in order to effectuate powers in actuality we need a specific actual set-up.

{\it A power is a contextual existent}. A power can produce an actual effectuation once the preconditions for its exposure become actual, i.e. once we construct in space-time the definite experimental arrangement which allow us to express the power. The potential existence of a power is not determined by the choice of an experimental procedure in the laboratory, it is only the actual effectuation of a power which needs the specific context in order to express itself. The choice of the context does not change the PSA, it only builds the bridge between the PSA, which contains a definite set of powers, and the actual effectuation of one of them in the ASA provided by the specific experimental set-up. Thus within our scheme, we recover the notion of objective measurement as exposing a preexistent state of affairs. The difference is that in QM, according to our interpretation, a measurement does not expose an ASA, as in classical physics, in QM a measurement exposes a PSA.

\section{`What can be Observed?' or `What can be Felt?'}

Our research regarding the possibility to put forward a metaphysical interpretation of QM has been guided by the words of Einstein: ``It is only the theory which can tell you what can be observed.'' But what is observed according to QM? Or in other words, what is a {\it quantum experience}? This is of course a very deep and difficult question which has remained neglected due to the presupposition ---forced again by the Danish physicist \cite{BokulichBokulich05}--- that experience must be restricted to classical space-time experience. 

{\it QM is a tactile theory.} QM is not a visual theory, one does not observe powers in the same way one is taught to see entities. In classical mechanics measurement takes place through {\it observation}. This is a completely objective process which allows us to examine and determine the properties of the entities under study. Observation appears from dissecting the change in the properties involved, in doing so we seek identities leaving aside everything which is not. Finally the properties are brought into a whole by the notion of entity. But in QM things take place in a very different way. Powers make themselves present in the actual realm through {\it action} and {\it change}. The quantum postulate ---which we have also called: {\bf principle of difference}--- is one of the basic cornerstones of the quantum experience within actuality. As noted by Bohr \cite[p. 53]{Bohr34}: ``[the essence of quantum theory] may be expressed in the so-called quantum postulate, which attributes to any atomic process an essential discontinuity, or rather individuality, completely foreign to the classical theories and symbolized by Plank's quantum of action.'' Indeed, one of the most important features of the quantum experience is that in QM we only measure shifts of energy, elementary processes, pure difference. As noticed by Nancy Cartwright:

\begin{quotation}
{\small ``It makes good sense to take energy transitions as basic for the interpretation of quantum mechanics. For it is only through interchanging energy that quantum systems interact and can register their interactions in a macroscopically observable way. In a very well-known argument against reduction of the wave packet, Hans Margenau has urged that all measurements are ultimately measurements of position. But this should be pushed one step further. All position measurements are ultimately measurements of energy transitions. No matter that a particle passes by a detector ---the detector will not register unless it exchanges some energy with the particle. The exchanges of energy is the basic event that happens in quantum mechanics; and the basic event whose effects are theoretically described and predicated.'' \cite[p. 55]{Cartwright78}}
\end{quotation}

Contrary to classical mechanics, which measures {\it identities} ---a one to one correlation between object and apparatus---, in QM we always measure {\it differences}. Experience through difference is a tactile experience. QM has imposed a different primacy to the modes of sensation, where the most important of all (according to classical physics) needs to be forgotten. QM is a land of darkness were sensation appears through discrete `imprints', `clicks' and `ticks'. In QM we do not see, {\it we touch}, we do not observe, {\it we feel} in a discrete manner through photographic plates, detectors and cloud chambers. Even light appears to be interesting only when regarded as a fog through which shadows can be seen. What is felt in QM through an {\it elementary processes} (shift of energy) is a power. By repeating the experience we can learn about the potentia of each one of the powers present in a given PSA ---just in the same way that in order to learn about an object, in a given ASA, we must observe it from multiple perspectives and angles. 

The elementary process produces the actual effectuation as an expression of a power which exists in the mode of being of potentiality. It is through a shift in energy, be that a click in a Geiger counter, a spot in a photographic plate or a small droplet in a cloud chamber, that the power gets exposed in space-time. The powers, which rest in the potential mode of existence, are thrown through the {\it immanent cause} into actuality when the elementary process takes place. However, the power still remains in the potential, regardless of actuality. Because of the immanent cause its existence remains always independent of its effectuation in the actual realm \cite{deRonde13b}.

It is the {\it immanent cause} which allows us to connect the power with its actual effectuation without destroying nor deteriorating the power itself. The immanent cause allows for the expression of effects remaining both in the effects and its cause. It does not only remain in itself in order to produce, but also, that which it produces stays within. Thus, in its production of effects the potential does not deteriorate by becoming actual ---as in the case of the hylomorphic scheme. Actual results are single effectuations, singularities which expose the superposition in the actual mode of existence, while superpositions remain evolving deterministically according to the Schr\"{o}dinger equation in the potential mode of existence, even interacting with other superpositions and producing new potential effectuations.

\section{Actual Effectuations and Potential Effectuations}

{\it Powers are objective existents.} Just like entities exist, even when there is no light to see them, powers exist in the world, even when we do not feel them as actualities in space-time. Powers exist even when there in no `click' or `tick' in a receptor, just like the moon is out there, whether or not we look at it. In the same way that entities appear to us through observation, and remain in the dark when light does not shine upon them, powers can be experienced through the actualization of elementary processes. In QM only change, shifts of energy must be taken into account, the quantum postulate implies a different way of acquiring sense data, a new experience which does not end exclusively in actuality and deserves further investigation. In the history of physics, due to a specific development of metaphysics, we have been accustomed to see, not to feel. We have learnt that physics is either about the observation of actualities (empiricism) or about entities which exist ---independently of observation--- in the mode of being of actuality constituting an ASA (realism). According to our theory of powers, there is more than actual effectuations. 

{\it Powers have both actual effectuations and potential effectuations.} An effect in space-time is what we call an {\it actual effectuation}. Analogously, our scheme proposes the existence of powers, each of which possesses a definite potentia and exists in the potential mode of being constituting a PSA. The interaction between powers is what we call a {\it potential effectuation}. Potential effectuations ---and not actual effectuations--- constitute the main type of interaction in QM and are governed by the Schr\"{o}dingier equation of motion ---in analogous way as actual interactions in classical mechanics are governed by the Newtonian equation of motion. Powers can have actual effectuations, which is the way we have learnt about QM in the first place. However, it is potential effectuations ---rather than actual ones--- which allow us to design quantum computers, quantum teleportation and all sorts of quantum information processing. By escaping from the limits imposed by actualism, we might be able to think anew not only many of the paradigmatic problems and paradoxes raised in QM, but also quantum experience itself. We will come back to this analysis in the final section of this paper.

\section{Representing Quantum Mechanics: Powers, Potentia and Potential Effectuations}

We are now ready to present our interpretation in axiomatic form. In this interpretation, which takes as a standpoint the orthodox formalism of QM, there are no systems, no states nor properties involved. Instead we have to provide the connection between the mathematical formalism and its physical interpretation provided via our new set of concepts: power, potentia, elementary process and potential effectuation. 

\begin{enumerate}

{\bf \item[I.] Hilbert Space:} QM is represented in a vector Hilbert space.

{\bf \item[II.] Potential State of Affairs (PSA):} A specific vector $\Psi$ with no given mathematical representation (basis) in Hilbert space represents a PSA; i. e., the definite existence of a multiplicity of powers, each one of them with a specific potentia.

{\bf \item[III.] Actual State of Affairs (ASA):} Given a PSA and the choice of a definite basis ${B, B', B'',...}$ (or equivalently a C.S.C.O.) a context is defined in which a set of powers, each one of them with a definite potentia, are univocally determined as related to a specific experimental arrangement (which in turn corresponds to a definite ASA). The context builds a bridge between the potential and the actual realms, between quantum powers and classical objects. The experimental arrangement (in the ASA) allow for the powers (in the PSA) to express themselves in actuality through elementary processes which produce actual effectuations.

{\bf \item[IV.] Faculties, Powers and Potentia:} Given a PSA, $\Psi$, and the context we 
call a faculty to any superposition of more than one power. In general given the basis $B= \{ | \alpha_i \rangle \}$ the faculty $F_{B, \Psi}$ is represented by the following superposition of powers:

\begin{equation}
c_{1} | \alpha_{1} \rangle + c_{2} | \alpha_{2} \rangle + ... + c_{n} | \alpha_{n} \rangle
\end{equation}

We write the faculty of the PSA, $\Psi$, in the context $B$ in terms of the order pair given by the elements of the basis and the coordinates of the PSA in that basis:

\begin{equation}
Faculty \ (B, \Psi) = (| \alpha_{i} \rangle, c_{i})
\end{equation}

The elements of the basis, $| \alpha_{i} \rangle$, are interpreted in terms of {\it powers}. The coordinates of the elements of the basis, $c_{i}$, are interpreted as the {\it potentia} of the power $| \alpha_{i} \rangle$, respectively. Given the PSA and the context the faculty $F_{B, \Psi}$ is univocally determined in terms of a set of powers and their respective potentia. 

Notice that in contradistinction with the notion of quantum state the definition of a faculty is basis dependent.

{\bf \item[V.] Elementary Process:} In QM one can observe discrete shifts of energy (quantum postulate). These discrete shifts are interpreted in terms of {\it elementary processes} which produce actual effectuations. An elementary process is the path which undertakes a power from the potential realm into its actual effectuation. This path is governed by the {\it immanent cause} which allows the power to remain preexistent in the potential realm independently of its actual effectuation. Each power $| \alpha_{i} \rangle$ is univocally related to an elementary process represented by the projection operator $P_{\alpha_{i}} = | \alpha_{i} \rangle \langle \alpha_{i} |$.

{\bf \item[VI.] Actual Effectuation of Powers (Measurement):} Powers exist in the mode of being of ontological potentiality. An {\it actual effectuation} is the expression of a specific power in actuality. Different actual effectuations expose the different powers of a given faculty. In order to learn about a specific PSA (constituted by a set of powers and their potentia) we must measure repeatedly the actual effectuations of each power exposed in the laboratory.

Notice that we consider a laboratory as constituted by the set of all possible experimental arrangements. 

{\bf \item[VII.] Potentia (Born Rule):} A {\it potentia} is the strength of a power to exist in the potential realm and to express itself in the actual realm. Given a PSA, the potentia is represented via the Born rule. The potentia $p_{i}$ of the power $| \alpha_{i} \rangle$ in the specific PSA, $\Psi$, is given by:

\begin{equation}
Potentia \ (| \alpha_{i} \rangle, \Psi) = \langle \Psi | P_{\alpha_{i}} | \Psi \rangle = Tr[P_{ \Psi} P_{\alpha_{i}}]
\end{equation}

In order to learn about a faculty we must observe not only its powers (which are exposed in actuality through actual effectuations) but we must also measure the potentia of each respective power. In order to measure the potentia of each power we need to expose the faculty statistically through a repeated series of observations. The potentia, given by the Born rule, coincides with the probability frequency of repeated measurements when the number of observations goes to infinity.

{\bf \item[VIII.]  Potential Effectuation of Powers (Schr\"odinger Evolution):} Given a PSA, $\Psi$, powers and potentia evolve deterministically, independently of actual effectuations, producing {\it potential effectuations} according to the following unitary transformation:

\begin{equation}
i \hbar \frac{d}{dt} | \Psi (t) \rangle = H | \Psi (t) \rangle
\end{equation}
\end{enumerate}

\section{Adequate Problems and the Deconstruction of Quantum Paradoxes}

As mentioned above, we believe it was Mach who, through his criticism to the {\it a priori} dogmatic concepts which governed Newtonian physics, was able to create the conditions of possibility for a new region of thought. Going against metaphysical dogmatism these new conditions inaugurated one of the most outstanding revolutionary periods in western thought. One of its results was the creation of two physical theories: QM and Relativity. But, as Constantin Piron has remarked, these were in the end ``two failed revolutions'' \cite{Piron99} for, up to the present, none of these theories has been truly understood. In particular, we believe that the failure of the quantum revolution stands on the fact that we have not been able to create the concepts which would allow us to think radically in terms of this theory. Instead, most attempts ---following Bohr's correspondence principle--- have concentrated their efforts in trying to justify the path from the quantum realm into classical mechanics. Indeed, Niels Bohr is maybe one of the most important figures responsible for the present state of affairs in physics. According to him ``[...] the unambiguous interpretation  of any measurement must be essentially framed in terms of classical physical theories, and we may say that in this sense the language of Newton and Maxwell will remain the language of physicists for all time.'' But more importantly, ``it would be a misconception to believe that the difficulties of the atomic theory may be evaded by eventually replacing the concepts of classical physics by new conceptual forms.'' \cite[p. 7]{WZ} His pragmatically based conception of physics together with his understanding of experience as `spacio-temporal experience' not only evaded the creation and development of new physical concepts but also precluded the interrogation regarding the meaning and understanding of QM itself. 

The definition of a problem configures in itself not only the possible questions to be made but also, and more importantly, the limits of the possible answers and solutions. This is why we have been interested in disclosing the stances and presuppositions involved in many of the orthodox problems discussed in the literature. Elsewhere \cite{deRonde11}, we engaged in the configuration of a map which could help us to disclose the hidden agenda of several paradigmatic problems in QM. According to this map there are two main problems in the interpretation of QM. The first problem is the so called ``basis problem'' which discusses how to justify the path from the set of multiple incompatible contexts to the actual experimental set up in the laboratory. As it is well known, contextuality is the main feature which avoids the consideration of the multiple basis as an ASA. This aspect of the theory is commonly addressed and understood in the literature as a problem that needs to be resolved, a character of QM that needs to be somehow bypassed or erased in order to explain and justify how, from the contextual formalism of QM, we arrive to a classical Boolean experimental arrangement. The second problem is the so called ``measurement problem'' which, given a quantum superposition of states, attempts to justify how at the end of the measurement we always obtain a single result ---instead of a superposition of them. Both problems presuppose as a solution our classical physical description of the world ---in terms of an ASA--- and try to fit QM {\it within}. Put in  a nutshell both problems attempt to justify the actual ``common sense'' mode of existence and experience. 

Contrary to this teleological line of research which seems to focus on the justification of a presupposed answer to the question of interpretation, we believe that the most important problem of QM regards the productive creation of understanding and representation. Our strategy has been to produce an {\it inversion} of both ``basis'' and ``measurement'' problems \cite{deRonde11}.  Such an inversion consists in subverting the problems and ask, firstly, about the meaning regarding the preexistence of the multiple contexts ---instead of trying to justify the actual existence of the classical measurement set-up---, and secondly, about the meaning and representation of a quantum superposition ---instead of trying to justify the single result we find at the end of a measurement. Instead of justifying actuality our approach has focused in trying to develop a potential realm which matches the formal requirements of QM. In this paper we have attempted to provide an answer the first question, which deals with the existence of the many incompatible contexts, through the introduction of a specific notion: the PSA. Such a preexistent mode of existence ---which escapes the limits of actuality--- allows us to recover an objective account of physical reality for QM escaping at the same time the problematic entanglement between subjective and objective aspects of the theory ---introduced by forcing the formalism into the ontology of entities. The second question, which deals more specifically with the conceptual representation of the quantum mechanical formalism and the meaning of superpositions, has been discussed through the development of the notions of `power' and `potentia'. One of the main consequences of this proposal regards the introduction of `potential effectuations', a notion which opens the door not only to the ontological consideration of many of present-day technical developments in QM, but also to the reconsideration of the meaning itself of {\it physical experience}.

\end{document}